\documentclass[12pt]{article}

\usepackage{graphicx} % pour l'inclusion de figures en eps,pdf,jpg
\usepackage{hyperref,url}
\usepackage{color}
\usepackage{amssymb,latexsym}
\usepackage{amsmath,amsthm}
\usepackage{enumerate}
\usepackage{tikz}
\usetikzlibrary{shapes,positioning}
\usepackage{latexsym,amsfonts,amscd,amssymb,verbatim,hyperref}
\usepackage{dsfont}
\usepackage[affil-it]{authblk} % affiliation auteurs

\usepackage{palatino}

\usepackage[applemac]{inputenc}
\usepackage[english]{babel}

\usepackage{geometry}
\usepackage{setspace} % Pour changer l'interligne

\usepackage[braket, qm]{qcircuit}

\theoremstyle{plain}
\newtheorem{theorem}{Theorem}%[section]

\theoremstyle{definition}

\newtheorem{definition-theorem}[theorem]{Definition-Theorem}
\newtheorem{example}[theorem]{Example}

\theoremstyle{remark}
\newtheorem{remark}{Remark}

\newcommand{\Uf}{\mathsf U} % Functional representation
\newcommand{\suchthat}{\colon} %such that
\newcommand{\bbR}{\mathbb R} %Reals

\newcommand{\bol}{\boldsymbol} %boldsymbol
\newcommand{\et}{~~\text{and}~~} % and (in condition)
\newcommand{\disand}{\qquad \text{and}\qquad} % and (display mode)
\newcommand{\one}{\mathds{1}}
\newcommand{\calO}{\mathcal{O}}

\newcommand{\tr}{\mathrm{tr}_}

\newcommand{\ketbra}[2]{| #1 \rangle \langle #2 |}

\newcommand{\cnot}{\text{Cnot}}
\newcommand{\notgate}{N}
\newcommand{\Hgate}{H}

\newcommand{\doublequote}[1]{``#1''}

\def \be {\begin{equation}}
\def \ee {\end{equation}}
\def \bes {\begin{equation*}}
\def \ees {\end{equation*}}
\def \baa {\begin{align}}
\def \eaa {\end{align}}
\def \baas {\begin{align*}}
\def \eaas {\end{align*}}
\def \bea {\begin{eqnarray}}
\def \eea {\end{eqnarray}}
\def \beas {\begin{eqnarray*}}
\def \eeas {\end{eqnarray*}}

\title{The ABC of Deutsch--Hayden Descriptors
 }
\author{Charles Alexandre B\'edard}
\affil{\small {Universit\`a della Svizzera italiana}\\
\footnotesize \emph{charles.alexandre.bedard@usi.ch}}
	
\begin{document}
\date{April 2021}					
\maketitle							
%\tableofcontents

\begin{abstract}
%It has been more than 20 years since Deutsch and Hayden introduced the formalism of descriptors, showing it to provide an entirely local account of quantum systems.
%
It has been more than 20 years since Deutsch and Hayden proved the locality of quantum theory, using the Heisenberg picture of quantum computational networks. 
%It has been more than 20 years since Deutsch and Hayden leveraged the Heisenberg picture to coin the formalism of descriptors, showing it to provide an entirely local account of quantum systems.
%
Of course, locality holds even in the face of entanglement and Bell's theorem.
Today, most researchers in quantum foundations are still convinced not only that a local description of quantum systems has not yet been provided, but that it cannot exist.
The main goal of this paper is to address this misconception by re-explaining the descriptor formalism in a hopefully accessible and self-contained way. 
It is a step-by-step guide to how and why descriptors work.
Finally, superdense coding is revisited in the light of descriptors.
\end{abstract}

\section{Motivation}\label{secintro}

It is still a widespread belief that a complete description of a composite entangled quantum system cannot be obtained by descriptions of the parts, if those are expressed independently of what happens to other parts. This apparently holistic feature of entangled quantum states entails 
violation of Bell inequalities~\cite{bell1964, aspect1982experimental} and quantum teleportation~\cite{bennett1993teleporting}, which are repeatedly invoked to sanctify the ``non-local'' character of
 quantum theory. But this widespread belief has been proven false more than twenty years ago by Deutsch and Hayden~\cite{deutsch2000information}, who by the same token provided an entirely local explanation of Bell-inequality violations and teleportation. 
  
Descriptions of dynamically isolated --- but possibly entangled --- systems~$A$ and~$B$ %and~$AB$
are \emph{local} if that of~$A$ is unaffected by any process system~$B$ may undergo, and vice versa. After Bell, it has become conventional wisdom to equate locality with a possible explanation by a local hidden variable theory. However, local hidden variables are only one way in which locality can be instantiated~\cite{brassard2019parallel}. Here, locality is taken in its crudest form, the one advocated by Einstein: ``the real factual situation of the system~$S_2$ is independent of what is done with the system~$S_1$, which is spatially separated from the former'' \cite{schilppalbert1970}.
Descriptions of individual systems~$A$ and~$B$ are \emph{complete} if, when put together, they can predict the distributions of any measurement performed on the whole system~$AB$.

For instance, if~$AB$ is in a pure entangled state~$\ket{\Psi}^{AB}$, the reduced density matrices
\bes
\rho^A= \tr{B} \ketbra{\Psi}{\Psi} \disand \rho^B = \tr{A} \ketbra{\Psi}{\Psi}
\ees
are local but incomplete descriptions. 
This is because~$\rho_A$ is left unaffected regardless of what happens to system~$B$, however, since~$\ket{\Psi}^{AB}$ is entangled, it or its associated density matrix~$\ketbra{\Psi}{\Psi}$ can no longer be recovered from~$\rho^A$ and~$\rho^B$. Some information that could reveal crucial to compute the distribution of some joint measurements has been discarded in the tracing out.
If instead the descriptions of~$A$ and~$B$ are both taken to be the global wave function~$\ket{\Psi}^{AB}$, then one finds a complete but non-local account. 

We seem to be stuck in a dichotomy, apparently forced to describe quantum systems either non-locally or incompletely. But the dichotomy is false. 
%
%In the spirit of Heisenberg's original formulation of quantum theory, in which dynamical variables are promoted to matrices,
Following Gottesman's~\cite{gottesman1999heisenberg} quantum computation in the Heisenberg picture, Deutsch and Hayden define so-called \emph{descriptors} for individual qubits and showed this mode of description to be both local and complete, hence vindicating the locality of quantum theory. 
In other words, even entangled systems admit a separable description. 
 When such a bold foundational result collects a mere 190 citations in more than 20 years, it is evidence that a large portion of the community of quantum foundations is unaware of the idea, or worse, does not understand it. 
This is the problem that this paper addresses, and it does so by providing a detailed and self-contained explanation of how and why descriptors work. The paper culminates with the superdense coding protocol being revisited in the established framework.
It is aimed both for experts and non-experts in quantum theory. A background in physics is optional; Only introductory knowledge in quantum information theory is required.

\section{A Question of Picture}

In quantum theory, computations leading to statistics of measurable quantities all take the same form, namely, that of Dirac's celebrated bra-ket notation,~$\bra{\cdots} \cdots \ket{\cdots}$.
Physicists recognize this kind of computation as the expected value of some observable. Quantum information scientists, bear with me for another 10 lines. An observable $\calO$ is represented by a hermitian operator which admits a spectral decomposition
\bes
\calO = \sum_i \lambda_i \Pi_i \,,
\ees
where $\lambda_i \in \bbR$ is an eigenvalue corresponding to the measurement outcome and $\Pi_i$ is the corresponding projector on the eigensubspace. If the system is in state~$\ket{\psi}$, the expected value of such an observable is given by~$\bra{\psi} \calO \ket{\psi}$, since
\bes \label{eqev}
\bra{\psi} \calO \ket{\psi} = \bra{\psi} \sum_i \lambda_i \Pi_i \ket{\psi} = \sum_i \bra{\psi} \Pi_i \ket{\psi} \lambda_i = \sum_i p_i \lambda_i 
  \,,
\ees
where $p_i$ can be thought of the probability of measuring outcome $\lambda_i$. 
While this type of computation is routine for physicists, quantum information scientists usually compute probabilities of measurement outcomes. An $n$-qubit network in the state 
\bes
\sum_{j=0}^{2^n-1} \alpha_j \ket j
\ees
has a probability $|\alpha_l|^2$ to return the classical value \doublequote{$l$}. But
\bes
|\alpha_l|^2 = \bra \psi \ketbra{l}{l} \ket \psi \,
\ees
is nothing but the expectation value of the observable $\ketbra{l}{l}$. Hence, the reader who is unfamiliar with observables can simply keep in mind projectors of the form $\ketbra ll$ required to compute probabilities, but this\footnote{
The most general observable can be thought as a choice of basis~$\{\ket{\phi_i} \}_i$, with a real number~$\lambda_i$ corresponding to each basis vector. Indeed, \mbox{$\calO = \sum_i \lambda_i \ketbra{\phi_i}{\phi_i}$} defines a generic hermitian operator, so a generic observable. Constructing an observable in this way makes ``measuring an observable'' clearer for quantum information scientists: It corresponds to measuring in the basis~$\{\ket{\phi_i} \}_i$ with the measurement outcomes labelled by~$\lambda_i$. Of course, this measurement can be done by performing the unitary that maps the basis~$\{\ket{\phi_i} \}_i$ to the computational basis~$\{ \ket i \}_i$, before measuring in this last basis.} footnote explains further. 

A generic state $\ket \psi$ arises from the evolution of an initial state that shall be denoted $\ket{0}$. If $U$ is the unitary operator representing this evolution,~\mbox{$\ket \psi = U \ket 0 $}, so the computations carried to predict measurable quantities all have the form
\be \label{eq:sand}
\bra 0 U^\dagger \calO U \ket 0 \,.
\ee

The \emph{Schr\"odinger picture} is about viewing the sandwich equation \eqref{eq:sand} as if the bread evolves and the meat stays constant, namely,
\bes
\left(\bra 0 U^\dagger \right) \, \calO \, \left(U\vphantom{^\dagger} \ket 0\right) \,.
\ees
With such a viewpoint, the initial state $\ket 0$ evolves to the final state \mbox{$\ket \psi= U\ket 0$} and the observable $\calO$ remains constant.

The \emph{Heisenberg picture} is about regarding the sandwich equation as if the meat evolves but the bread remains constant,
\bes \label{eq:bread}
\bra 0 \left( U^\dagger \calO U \right) \ket 0 \,.
\ees
In this picture, the state vector remains fixed to $\ket 0$ but the observable $\calO$ evolves to $ U^\dagger \calO U $. Therefore, in the Heisenberg picture, the term `state', which refers to a quantity that is fixed to $\ket 0$, becomes a misnomer. It will thus be called the \emph{reference vector}.
But then, in the Heisenberg picture, can the quantum information of the system at a given time be encoded in a single mathematical object? Yes: It is precisely what the \emph{descriptor} does.

\section{Tracking Observables}

In the Heisenberg picture, a quantum system shall no longer be described by its state vector, but rather by an object that encodes the information about \emph{all} the evolved observables of the system. 
This is a tall order since there is an uncountable number of such observables.
 Things are greatly simplified once it is realized that observables are linear operators and that the latter form a vector space. Since the evolution $\calO \to U^\dagger \calO U$ is linear, one does not need to track the evolution of infinitely many observables: \emph{Only a basis} of the linear operators suffices. Indeed, if $\calO = \sum_j a_j B_j$, then $U^\dagger \calO U = \sum_j a_j U^\dagger B_j U$, so it suffices to track how each operator $B_j$ of the basis evolves by $U$ to then compute how any observable evolves.

\subsection*{The Descriptor of a $1$-Qubit Network}

In the case of a singe qubit, the Pauli matrices together with the identity,
\bes
\bol \sigma = (\sigma_x, \sigma_y, \sigma_z) = \left( \begin{bmatrix} 0 & 1 \\ 1 & 0 \end{bmatrix},
\begin{bmatrix} 0 & -i \\ i & 0 \end{bmatrix},
\begin{bmatrix} 1 & 0 \\ 0 & -1 \end{bmatrix} \right) ~\et~ \sigma_0 = \one = \begin{bmatrix} 1 & 0 \\ 0 & 1 \end{bmatrix}
\ees
form a basis of the $2\times2$ matrices, if the linear combinations are taken over complex numbers. Following the evolution of $\one$ is trivial, \mbox{$U^\dagger \one U = \one$}, so it can be neglected. This means that one only needs to follow the evolution of $\boldsymbol \sigma$, to then be able to recover any evolved observable, or the expectation value thereof. 

Hence, for a single qubit quantum network, the \emph{descriptor} of the qubit at time $t$ is given by
$$\bol q(t) = U^\dagger \boldsymbol \sigma U\,,$$
where $U$ is the unitary operator that represents the evolution undergone by the quantum network between time~$0$ and time~$t$.

\begin{example} \label{ex:H}
Consider the following quantum circuit 
\begin{center}
\begin{picture}(80,35)(-40,-7)
\put(-40,0){\makebox(20,20){$\ket 0$}}
\put(-20,10){\line(1,0){20}}
\put(0,0){\framebox(20,20){$H$}}
\put(20,10){\line(1,0){20}} %ligne pointillée
\multiput(-10,-2)(0,2){13}
{\line(0,1){1}}
\put(-19,-11){\footnotesize{$t=0$}}
\put(20,10){\line(1,0){20}}
\multiput(30,-2)(0,2){13}
{\line(0,1){1}}
\put(21,-11){\footnotesize{$t=1$}}
\end{picture} \hspace{20pt} \raisebox{10 pt}{where~$\qquad H=  \frac{1}{\sqrt 2}  \begin{bmatrix} 1&1\\1&-1
\end{bmatrix}$}
\end{center}
is the Hadamard gate. At time $t=0$, the descriptor is~\mbox{$\bol q(0)= \bol \sigma = (\sigma_x, \sigma_y, \sigma_z)$}, while at time $t=1$, the descriptor is
\be \label{eq:ex1}
\bol q(1) = H^\dagger \bol \sigma H = H^\dagger(\sigma_x,\sigma_y, \sigma_z)H = (\sigma_z, -\sigma_y, \sigma_x)\,.
\ee
The Heisenberg picture and the expression for~$\bol q(1)$ can be used to compute the probability of measuring the outcome ``$0$''. Representing $\ket 0$ and $\ket 1$ respectively by $ \begin{bmatrix} 1 \\  0 \end{bmatrix}$ and $ \begin{bmatrix} 0 \\  1 \end{bmatrix}$,

\beas
\bra 0 H^\dagger \ketbra 00 H \ket 0
&=& \bra 0 H^\dagger \begin{bmatrix} 1 & 0 \\ 0 & 0 \end{bmatrix} H \ket 0 \\
&=& \bra 0 H^\dagger \left ( \frac{\one + \sigma_z}{2} \right) H \ket 0 \\
&=&  \bra 0 H^\dagger \left ( \frac{\one + q_z(0)}{2} \right )H \ket 0 \\
&=& \bra 0 \frac{\one + q_z(1)}{2} \ket 0 \\
&=& \frac{\bra 0 \one \ket 0  + \bra 0 \sigma_x \ket 0}{2} \\
&=& \frac 12 \,.
\eeas

\end{example}

\subsection*{Descriptors of an $n$-Qubit Network}

Consider now, and for the rest of the paper, the case of $n$ interacting qubits in a quantum computational network. Suppose that the qubits are initialized at time~$0$ in the state~$\ket 0^{\otimes n}$, which, when more conveniently denoted $\ket 0$, correspond to the Heisenberg reference vector so far invoked. Although this network seems like a restricted system, its ability to simulate any other quantum system to arbitrary accuracy~\cite{deutsch1989quantum} makes it completely general.  
Moreover, no generality is lost by assuming that each gate in the network requires exactly one unit of time, so that the state of the network needs only be specified at integer values of time. Let again~$U$ be the unitary operator representing the evolution of the network between time~$0$ and time~$t$.

A natural basis of the space of all operators on $n$ qubits is the product of Pauli operators, namely,
\bes \label{eq:myset}
 \mathcal B \equiv \left \{ \sigma_{\mu_1} \otimes  \sigma_{\mu_2} \otimes \ldots  \sigma_{\mu_n} \, \colon \, \mu_i \in \{ 0,x,y,z\} \right \} \,.
\ees
There are $4^n$ such matrices, and they are linearly independent, so, indeed, they form a basis of the $2^n \times 2^n = 4^n$ dimensional complex\footnote{In fact,~$\mathcal B$ is a basis of hermitian operators if~\emph{real} linear combinations are considered. However, in the present context, it is more relevant to think of~$\mathcal B$ as a basis of all linear operators.} vector space of linear operators on $n$-qubits.

This means that if one knows how each observable of the basis $\mathcal B$ evolves by the action of~$U$,
\bes
 \sigma_{\mu_1} \otimes  \sigma_{\mu_2} \otimes \ldots  \sigma_{\mu_n}   \to U^\dagger
 \sigma_{\mu_1} \otimes  \sigma_{\mu_2} \otimes \ldots  \sigma_{\mu_n} U \,, \qquad \mu_i \in \{ 0,x,y,z\} \,,
\ees
then one knows, by linearity, how each observable evolves.

\subsection*{The Main Simplification}

A great simplification is to track the evolution of only the set of observables
\be \label{eq:q0}
\bol q_i(0) =  \one^{i-1} \otimes \boldsymbol \sigma \otimes \one^{n-i}\,, \qquad i= 1, \ldots , n \,,
\ee
where $\one^k$ stands for the tensor product of $k$ copies of the identity. Note that for each~$i$,~$\bol q_i(0)$ has 3 components, each of them being an operator acting on the whole Hilbert space.
The $n$-tuple whose components are the~$\bol q_i(0)$ is denoted~$\bol q(0)$. Bold quantities are vectors, so for instance one writes~$\bol q_i(0)$, but~$ q_{ix}(0)$. 
This~$\bol q_i(0)$ is \emph{the descriptor of qubit~$i$ at time~$0$}. The descriptor at time~$t$ is then given by
\be \label{eq:qt}%\tag{EVO 1}
\bol q_i(t) = U^\dagger \bol q_i(0) U  \,.
\ee

Importantly, note that $\bol q(0)$ contains many fewer components than $\mathcal B$ contains elements. In fact, instead of tracking the $4^n$ operators of~$\mathcal B$, only $3n$ are suggested here.
The reason is that these~$3n$ operators can be multiplied to generate any of the $4^n$ basis operators.
Moreover, this multiplicative structure is preserved by the evolution~$U$, namely, if an observable is generated multiplicatively by~$q_{iw}(0)q_{jw'}(0)$, then the evolved observable is given by
 \bes
 U^\dagger q_{iw}(0)q_{jw'}(0) U = U^\dagger q_{iw}(0) U U^\dagger q_{jw'}(0)U = q_{iw}(t)q_{jw'}(t) \,.
\ees
This observation obviously extends to larger products, as well as to sums of products of components of~$\bol q(0)$.

\begin{example}
Considering a 2-qubit network, the observable~$\ketbra{01}{01}$ can be expanded in the basis~$\mathcal B= \left \{\sigma_\mu \otimes \sigma_\nu \suchthat \mu, \nu \in \{0,x,y,z\} \right \}$, and then expressed in terms of $\bol q_1(0)$ and~$\bol q_2(0)$. Indeed,
\beas
\ketbra{01}{01} &=& (\ketbra{0}{0} \otimes \one) (\one \otimes \ketbra{1}{1}) \\
&=& \left( \frac{\one + \sigma_z}{2} \otimes \one \right)  \left( \one \otimes \frac{\one - \sigma_z}{2} \right )\\
&=& \frac 14 \left( \one^2 - \one \otimes \sigma_z + \sigma_z \otimes \one - \sigma_z \otimes \sigma_z \right ) \\
&=& \frac 14 \left( \one^2 - q_{2z}(0) + q_{1z}(0) - q_{1z}(0)q_{2z}(0) \right ) \,.
\eeas

This can then be used to express in terms of~$\bol q(t)$ the time-evolved counter-part of the observable, $U^\dagger \ketbra{01}{01} U$, under a an evolution~$U$ between time~$0$ and~$t$:
\beas
U^\dagger \ketbra{01}{01} U &=& \frac 14 \left( U^\dagger \one^2 U - U^\dagger q_{2z}(0) U + U^\dagger q_{1z}(0) U - U^\dagger q_{1z}(0) U U^\dagger q_{2z}(0) U \right ) \\
&=& \frac 14 \left( \one^2 - q_{2z}(t) + q_{1z}(t) - q_{1z}(t)q_{2z}(t) \right ) \,.
\eeas
\end{example}

\subsection*{The Algebra of Descriptors}

The addition and multiplication of components of descriptors grant them with an algebraic structure.
\begin{remark}
The operators of $\bol q(0)$ satisfy %\footnote{There is an abuse of language here. Technically, the~$\mathfrak{su}(2)^{\otimes n}$ algebra is the vector space generated by real linear combinations of products of elements of~$\bol q(0)$ (so the space of hermitian matrices~$2^n \times 2^n$). Hence, the components of~$\bol q(0)$ satisfy the algebraic relations of the~\emph{generators} of~$\mathfrak{su}(2)^{\otimes n}$. \com{The existence of this footnote depends on the mathematical rigour of the journal.} }
the $\mathfrak{su}(2)^{\otimes n}$ algebra, namely
\beas \label{eq:su2n}
[q_{iw} (0), q_{jw'}(0)] &=& 0\hphantom{q_{z}(0)} \qquad (i \neq j \text{ and } \forall w, w') \nonumber \\
q_{ix}(0) q_{iy}(0) &=& i q_{iz}(0)  \qquad(\text{and cyclic permutations})  \\
q_{iw}(0)^2 &=& \one\hphantom{q_{z}(0)} \qquad( \forall w)\,. \nonumber
\eeas
\end{remark}
In the first line, the bracket denotes the commutator, $[A,B] = AB - BA$.
The above algebraic relations follow from those of the Pauli matrices and %\footnote{In fact the algebra~$\mathfrak{su}(2)^{\otimes n}$ is \emph{a priori} abstract, defined solely by the relations among its elements. In this case, the tensor product of Pauli operators with identities are said to~\emph{represent} the algebra, providing a concrete way of thinking of the abstract elements in terms of matrices.},
from the factorized form of the descriptors at time~$0$, displayed in equation~\eqref{eq:q0}. After evolving by~$U$, the descriptors~$\bol q_i(t)$ shall in general loose their direct connection with Pauli matrices, as well as their factorized form, but still, they preserve their algebraic relations.

\begin{remark} \label{remt}
For any~$t$, $\bol q(t)$ satisfies the $\mathfrak{su}(2)^{\otimes n}$ algebra :
\beas
[ q_{iw}(t), q_{jw'}(t)] &=& q_{iw}(t) q_{jw'}(t) - q_{jw'}(t) q_{iw}(t) \\
&=& U^\dagger q_{iw}(0) U  U^\dagger q_{jw'}(0) U  - U^\dagger q_{jw'}(0) U U^\dagger q_{iw}(0) U \\
&=& U^\dagger q_{iw}(0)  q_{jw'}(0) U  - U^\dagger q_{jw'}(0)  q_{iw}(0) U \\
&=& U^\dagger [q_{iw}(0), q_{jw'}(0)]  U \\
&=& 0 \,\hspace{16ex} (i \neq j \text{ and } \forall w, w') \\
&&\vphantom{ [ } \\
q_{ix}(t) q_{iy}(t) &=& U^\dagger q_{ix}(0) U U^\dagger q_{iy}(0) U \\
&=& U^\dagger q_{ix}(0)  q_{iy}(0) U \\
&=& U^\dagger i q_{iz}(0)  U \\
&=&  i q_{iz}(t)  \hspace{12ex} (\text{and cyclic permutations})\\
&&\vphantom{ [ } \\
q_{iw}(t)^2 &=& U^\dagger q_{iw}(0) U U^\dagger q_{iw}(0) U \\
&=& U^\dagger q_{iw}(0)  q_{iw}(0) U \\
&=& U^\dagger \one U \\
&=& \one \hspace{16ex} ( \forall w)\,. \hspace{15ex}
\eeas
\end{remark}

One might object that unitary evolution is but a special case of a larger class of processes represented by
\emph{completely positive and trace preserving} maps. Such processes include for instance noisy channels or maps that do not preserve the dimensionality of the system (and hence do not preserve the system's algebra). These processes are, however, a special case of unitary evolution. In fact, not only that, by Stinespring dilation theorem, these processes can be \emph{mathematically} understood as sub-processes of a larger unitary evolution, but they \emph{physically} are. Real quantum processes are unitary evolutions.

\subsection*{One More Simplification}

Following Gottesman~\cite{gottesman1999heisenberg}, the generating tuple $\bol q(0)$ could be reduced to $2n$ elements by noticing a redundancy due to the $\mathfrak{su}(2)^{\otimes n}$ algebra. In fact, for any $i$, only two of the triplet of operators $(q_{ix}(0), q_{iy}(0), q_{iz}(0))$ are required, since the omitted operator can be recovered by the product of the selected two. In what follows, the notation will not be modified, but one will happily use this shortcut to avoid tracking the observables~$q_{iy}(t)$, keeping in mind that~\mbox{$q_{iy}(t)= -i q_{ix}(t)q_{iz}(t)$}.

Summing this up, the Heisenberg picture is about tracking the evolution $\calO \to U^\dagger \calO U$ of uncountably many initial observables~$\calO$. This can be done by instead tracking the evolution~$\bol q(0) \to \bol q(t) = U^\dagger \bol q(0) U$ of only~$2n$ observables ($q_{iy}$ is omitted).
In fact,~$\bol q(t)$ allows to infer, by multiplication, the evolution of the $4^n$ observables of $\mathcal B$, which allow to infer, by linearity, the evolution of any observable. 

\section{Evolution from the Future?!}\label{subsec:wayout}

Although $\bol q(0) \to \bol q(t) = U^\dagger \bol q(0) U$
looks like a completely fine way in which observables should evolve, when $U$ is broken down into different gates, for instance \mbox{$U= G_t \dots G_2 G_1$}, one finds that the observables of the descriptors evolve in the wrong order! 
In fact, the order in which the gates are applied is first $G_1$, then $G_2$, and so on, until the last gate $G_t$ is applied.
However, the descriptors evolve as  
\be \label{eq:prob}
\bol q(0) \to G_1^\dagger G_2^\dagger \dots G_t^\dagger \bol q(0) G_t \dots G_2 G_1 \,.
\ee
The evolution of observables appears to occur from the last gate of the network to the first, which is inconvenient, since the network needs to be final before one can start to compute anything. Much worse, it does not reflect the actual dynamics that the system is undergoing, so this kind of evolution from the future simply cannot be the right explanation.

The way out of this conundrum is to notice that inasmuch as observables are linear operators generated by some set~$\bol q(0)$ of operators, the evolution operators --- or gates --- are too. They are generated multiplicatively and additively by the same set~$\bol q(0)$, since questions of hermiticity versus unitarity do not arise.
\subsection*{The Functional Representation of a Gate}
For a fixed gate with matrix representation~$G$, its multiplicative and additive generation by $\bol q(0)$ defines a function~$\Uf_{G}(\cdot)$ through
\bes \label{eqgateq0}
G=\Uf_{G}(\bol q(0))\,.
\ees 
The function~$\Uf_G(\cdot)$ takes value in unitary operators and will be referred to as the~\emph{functional representation} of the gate~$G$. Its functionality encodes the multiplicative and linear generation of~$G$ by the elements of~$\bol q(0)$. 
In other words, any matrix~$G$ can be expressed as a polynomial in the $2n$ matrices $q_{1x}(0), q_{1z}(0), \dots, q_{nz}(0)$, and~$\Uf_G(\cdot)$ is one such polynomial.
Now, when $\bol q(t)$ varies with $t$, the matrix representation $\Uf_G(\bol q(t))$ varies accordingly, but as we shall see in the next section, it is the fixed functionality of~$\Uf_G$ that plays a central algebraic role when performing computations in the Heisenberg picture.

\begin{example}
In the case of a single qubit network, the negation and Hadamard gates are described by
\beas
\notgate = 
\begin{bmatrix} 
0 & 1\\
1 & 0 
\end{bmatrix}
= \sigma_x = q_{x}(0)
\qquad \text{and} \qquad 
\Hgate = \frac{1}{\sqrt 2}
\begin{bmatrix} 
1 & 1\\
1 & -1 
\end{bmatrix}
= \frac{q_x(0) + q_z(0)}{\sqrt 2} \,,
\eeas
so their functional representations are 
\bes
\Uf_N(\bol q(t)) = q_x(t) \disand \Uf_H(\bol q(t))= \frac{q_x(t) + q_z(t)}{\sqrt 2} \,.
\ees
The counterclockwise rotation of a state vector in the $\ket 0$ \& $\ket 1$ plane\footnote{
Note that this operation represents the rotation of a polarized photon, but not exactly that of the spin of an electron. The reason for this is that a $\pi / 2$ rotation of a photon takes the horizontal polarization $\ket \leftrightarrow \equiv \ket 0$ to the vertical polarization $\ket{\updownarrow} \equiv \ket{1}$. However, the spin of an electron needs a~$\pi$ rotation to take the $\ket{\uparrow_z} \equiv \ket 0 $ to $\ket{\downarrow_z} \equiv \ket 1$.}
 is described by
\beas
R_\theta = \begin{bmatrix} \cos \theta & - \sin \theta \\  \sin \theta & \cos \theta \end{bmatrix} = \cos \theta~ \one - i \sin \theta~ \sigma_y 
= \cos \theta~ \one + \sin \theta~ q_{x}(0)q_{z}(0) \,,
\eeas
which defines its functional representation~$\Uf_{R_\theta}(\cdot)$.

In the case of an~$n$-qubit network, if such a unary gate, say~$H$, is applied on qubit~$i$, while all other qubits are left invariant, then the matrix representation of the corresponding evolution operator is
$$
H_i \equiv \one^{i-1} \otimes H \otimes \one^{n-i} = \frac{q_{ix}(0) + q_{iz}(0)}{\sqrt 2} \,,
$$
so its corresponding functional representation is $\Uf_{H_i}(\bol q(t))= \frac{q_{ix}(t) + q_{iz}(t)}{\sqrt 2}$.

\end{example}

\subsection*{Back in order!}

The apparently reversed-ordered evolution of equation~\eqref{eq:prob} can then be transformed back in the right order. 
Denoting \mbox{$V=G_{t-1}\dots G_2 G_1$}, one finds
\beas
\bol q (0) \to V^\dagger G_t^\dagger \bol q (0) G_t V 
&=& V^\dagger \Uf_{G_t}^\dagger(\bol q(0) ) \bol q (0) \Uf_{G_t}(\bol q(0) ) V \\
&=& V^\dagger \Uf_{G_t}^\dagger(\bol q(0) ) V V^\dagger \bol q (0) V V^\dagger \Uf_{G_t}(\bol q(0) ) V \\
&=&  \Uf_{G_t}^\dagger \left(V^\dagger \bol q(0) V  \right)~ V^\dagger \bol q (0) V~  \Uf_{G_t}\left(V^\dagger \bol q(0) V\right)  \\
&=&  \Uf_{G_t}^\dagger( \bol q(t-1))  V^\dagger \bol q (0) V  \Uf_{G_t}( \bol q(t-1) ) \,.
\eeas

In the second last line, the function $\Uf_{G_t}$ (and its hermitian conjugate) is applied to the components of $\bol q(0)$ that are sandwiched by $V^\dagger$ and $V$. The equality holds because if $\Uf_{G_t}$ contains products of components of $\bol q(0)$, the inner $V^\dagger$ and $V$ in the expansion of~$\Uf_{G_t}\left(V^\dagger \bol q(0) V\right)$ shall cancel out, leaving only the outer ones, which can then be factored out to retrieve the line before.

At this stage, the computation can be continued in two different ways. First, remembering that $V=G_{t-1}\dots G_2 G_1$, the argument can be iterated on both sides of the equation. This makes explicit that the problem of the order in which the observables evolve in the Heisenberg picture is solved by introducing the functional representation of the gates. Indeed, evolving the observables by the matrix representation of the gates acting in the wrong order,
\bes
G_1^\dagger G_2^\dagger \dots G_t^\dagger \bol q (0) G_t \dots G_2 G_1 \,,
\ees
is equivalent to the right ordering of the functional representation of the gates evaluated at the corresponding times, \emph{i.e.},
\bes
\Uf_{G_t}^\dagger (\bol q(t-1)) \dots \Uf_{G_2}^\dagger (\bol q(1)) \Uf_{G_1}^\dagger (\bol q(0)) ~ \bol q (0)~ \Uf_{G_1} (\bol q(0)) \Uf_{G_2} (\bol q(1)) \ldots \Uf_{G_t} (\bol q(t-1))\,.
\ees

Another way to continue the previous calculation is to invoke equation~\eqref{eq:qt} on both sides of the equation to find
\be \label{eq:qt2}
\bol q(t) = \Uf_{G_t}^\dagger( \bol q(t-1)) \bol q (t-1)  \Uf_{G_t}( \bol q(t-1) ) \,.
\ee
This is the way in which descriptors are prescribed to evolve in Ref.~\cite{deutsch2000information}. It is in fact correct and equivalent to equation \eqref{eq:qt}, although not trivially recognized.

\section{The Action on Desriptors}\label{sub:action}

Evolving the descriptor in a step-by-step fashion, as prescribed by equation~\eqref{eq:qt2}, permits to find out how a specific gate affects the different descriptors,~\emph{i.e.}, the action of the gate on the descriptors.
A gate $G_t$ transforms the~$2n$ components of~$\bol q(t-1)$ in the following way:
\bes
G_t \colon q_{iw}(t-1) \to q_{iw}(t) = \Uf_{G_t}^\dagger(\bol q(t-1) ) q_{iw}(t-1) \Uf_{G_t}(\bol q(t-1)) \,.
\ees
 Leveraging the fact that the descriptors at time~$t-1$ satisfy the~$\mathfrak{su}(2)^{\otimes n}$ algebra (\emph{c.f.} Remark~\ref{remt}), the functional representation~\mbox{$\Uf_{G_t}(\bol q(t-1))$} can be expanded and the algebraic relations of the many components of~$\bol q(t-1)$ that shall crop up are used to simplify the expression. As it shall be seen, the locality of the applied gate renders trivial most of those~$2n$ computations.

\begin{example}
Between time~$t-1$ and~$t$, let a Hadamard gate~$H$ be performed on the $i$-th qubit, so $G_t = H_i$. What is the action of~$H_i$ on~$\bol q_i$? And on~$\bol q_j$, with $j\neq i$? 

Recalling that
\bes
\Uf_{H_i}(\bol q(t-1)) = \frac{q_{ix}(t-1)+q_{iz}(t-1)}{\sqrt 2} \,,
\ees
the action on descriptor~$\bol q_i$ is then
\beas
 H_i \,\colon \, (q_{ix}(t-1), q_{iz}(t-1)) &\to& (q_{ix}(t), q_{iz}(t))\\
 &=& \Uf^\dagger_{H_i}(\bol q(t-1))  (q_{ix}(t-1), q_{iz}(t-1))  \Uf_{H_i}(\bol q(t-1))  \\
&=&  \frac{q_{ix}+q_{iz}}{\sqrt 2} (q_{ix}, q_{iz}) \frac{q_{ix}+q_{iz}}{\sqrt 2} \\
&=& \frac 12 (q_{ix}+q_{iz}+q_{iz}-q_{ix}, -q_{iz}+q_{ix}+q_{ix}+q_{iz})\\
&=& (q_{iz}, q_{ix}) \,.
\eeas
When the context does not require it, the time labels can be omitted, like here, from the third line onwards, the ``$(t-1)$'' has been discarded. One can then simply denote the action of the gate on the descriptors as~$H_{i} \,\colon \, (q_{ix}, q_{iz}) \to (q_{iz}, q_{ix}) $ without insisting on the time labels, since the calculation relies only on the time-independent algebra of descriptors. Notice that the result is analogous to what has been computed in Example~\ref{ex:H}, equation~\eqref{eq:ex1}, but here, no matrix multiplication was involved, only the algebra of descriptors. More specifically, the properties~$q_{iw}^2 = \one$ and~$q_{iz}q_{ix} = i q_{iy} = -q_{ix}q_{iz}$ have been used. 

How about the action of~$H_i$ on all other~$\bol q_j$, with $j \neq i$? Since~$\Uf_{H_i}(\bol q)$ depends only on $q_{ix}$ and~$q_{iz}$ (time labels removed), it commutes with~$\bol q_j$, leaving it invariant,
$$
\Uf^\dagger_{H_i}(\bol q)  (q_{jx}, q_{jz})  \Uf_{H_i}(\bol q) = \Uf^\dagger_{H_i}(\bol q) \Uf_{H_i}(\bol q)  (q_{jx}, q_{jz})  = (q_{jx}, q_{jz}) \,.
$$
\end{example} 

\subsection*{Locality and Completeness}

The fact that $\Uf_{H_i}$ depends only on $\bol q_i$ --- and so leaves invariant the descriptor of all qubits but qubit~$i$ --- is precisely due to the fact that $H_i$ is a gate that acts only on qubit~$i$. More generally, if the gate~$G_t$ acts only on qubits of the subset $I\subseteq \{1,2,\ldots,n\}$, then its functional representation~$\Uf_{G_t}$ shall only depend on components of~$\bol q_k(t-1)$, for~$k \in I$. For~$j \notin I$, the descriptor~$\bol q_{j}(t-1)$ shall then commute with~$\Uf_{G_t}(\bol q(t-1))$, so it will remain unchanged between times~$t-1$ and~$t$. Hence, anything that is done to any system that does not concern qubit~$j$ leaves its descriptor invariant, namely, \emph{the descriptors are a local description of quantum systems.}

The descriptors are also \emph{complete}, in that the expectation value of any time-evolved observable~$U^\dagger \calO U$ that concerns only qubits of~$I$ can be determined by the descriptors~$\bol q_k(t)$, with~$k\in I$.
This can be seen more clearly at time~$0$, where an observable~$\calO$ on the qubits of~$I$ is a linear (hermitian) operator that acts non-trivially \emph{only} on the qubits of~$I$. Any such operator can be generated additively and multiplicatively by the components of~$\bol q_k(0)$, with~$k\in I$, thereby defining a polynomial~$f_\calO(\cdot)$ for which
\bes
\calO = f_{\calO}(\{\bol q_k(0)\}_{k\in I}) \,, \qquad \text{and so} \qquad
U^\dagger \calO U = f_{\calO}(\{\bol q_k(t)\}_{k\in I})\,.
\ees

\begin{example}
Determine the action of~$N = \sigma_x$ and of~$\sigma_z$ on the descriptor of the qubit that is acted upon.
\beas
 N \,\colon \, (q_{x}(t-1), q_{z}(t-1)) &\to& (q_{x}(t), q_{z}(t))\\
 &=& \Uf^\dagger_{N}(\bol q(t-1))  (q_{x}(t-1), q_{z}(t-1))  \Uf_{N}(\bol q(t-1))  \\
&=&  q_{x} (q_{x}, q_{z}) q_{x} \\
&=& (q_{x}, - q_{z}) \,.
\eeas
%Here, the algebraic properties~$q_{ix}^2 = \one$ and~$q_{iz}q_{ix} = i q_{iy} = -q_{ix}q_{iz}$ have been used. 
Similarly, and with a lighter, time-independent notation,
\beas
 \sigma_z \,\colon \, (q_{x}, q_{z}) &\to& 
 \Uf^\dagger_{\sigma_z}(\bol q)  (q_{x}, q_{z})  \Uf_{\sigma_z}(\bol q)  \\
&=&  q_{z} (q_{x}, q_{z}) q_{z} \\
&=& ( - q_{x},  q_{z}) \,.
\eeas
\end{example}

\subsection*{The $\cnot$}

The controlled not gate, denoted~$\cnot$, is a two qubit gate of great importance. Not only does it represent a perfect measurement, but when the~$\cnot$ is supplemented by arbitrary unary gates, it forms a universal gate set. This means that any unitary transformation can be realized by a circuit with gates chosen solely among this set.

Consider a $\cnot$ gate where the qubit $c$ controls the target qubit $t$. Restricting to the subspace acted upon, the linear transformation is represented by
\bes
\cnot = \begin{bmatrix}
1&0&0&0\\
0&1&0&0\\
0&0&0&1\\
0&0&1&0
\end{bmatrix} \,.
\ees
The functional representation~$\Uf_{\cnot}(\cdot)$ is established by expressing the above matrix in terms of the components of the descriptor at time~$0$,
\beas
\cnot
&=&
\begin{bmatrix}
1&0&0&0\\
0&1&0&0\\
0&0&0&0\\
0&0&0&0
\end{bmatrix}
+
\begin{bmatrix}
0&0&0&0\\
0&0&0&0\\
0&0&0&1\\
0&0&1&0
\end{bmatrix} \\
&=& \vphantom{\int^A }\ketbra 00 \otimes \one + \ketbra 11 \otimes N \\
&=& \frac{\one + \sigma_z}{2} \otimes \one +  \frac{\one - \sigma_z}{2} \otimes \sigma_x  \\
&=& \frac 12 \left ( \one^2 + q_{cz}(0) + q_{tx}(0) - q_{cz}(0)q_{tx}(0) \right) \,.
\eeas
The functional representation of $\cnot$ ($c$ controls $t$) is thus given by
\bes
\Uf_{\cnot}(\bol q (t)) = \frac 12 (\one + q_{cz}(t) + q_{tx}(t) - q_{cz}(t)q_{tx}(t) ) \,.
\ees
The action of the~$\cnot$ on the descriptors that it affects can be found to be
\beas
\cnot \, \colon \, \left \{ 
		\hspace{-5pt}
		\begin{array}{c}
			(q_{cx} , q_{cz}) \\
			 (q_{tx} , q_{tz}) 
		\end{array}
		\hspace{-5pt}
	\right \}
\to
	\left \{
		\begin{array}{c}
		\hspace{-5pt}
			(q_{cx}  q_{tx}, q_{cz}) \\
			 (q_{tx} , q_{cz} q_{tz}) 
		\end{array}
		\hspace{-5pt}
	\right \} \,.
\eeas
For example, the calculation of $q_{cx} (t)$ is done below.
\beas
q_{cx}(t-1) &\to& q_{cx}(t) \\
&=& \Uf^\dagger_{\cnot}(\bol q (t-1))  q_{cx}(t-1)\Uf_{\cnot}(\bol q (t-1)) \\
&=&  \frac 14 \left ( \one + q_{cz} + q_{tx} - q_{cz}q_{tx} \right ) q_{cx}  \left (\one + q_{cz} + q_{tx} - q_{cz}q_{tx} \right) \\
&=& \frac 14 ( 
q_{cx} + q_{cx}  q_{cz} + q_{cx} q_{tx} - q_{cx} q_{cz} q_{tx} \\
&& \hspace{15pt}+ q_{cz} q_{cx}  + q_{cz} q_{cx} q_{cz} + q_{cz} q_{cx} q_{tx} - q_{cz} q_{cx} q_{cz} q_{tx}  \\
&& \hspace{15pt}+ q_{tx} q_{cx}  + q_{tx} q_{cx} q_{cz} + q_{tx} q_{cx} q_{tx} - q_{tx} q_{cx} q_{cz} q_{tx}  \\
&& \hspace{15pt}-q_{cz} q_{tx} q_{cx}  -q_{cz} q_{tx} q_{cx} q_{cz}  -q_{cz}q_{tx} q_{cx} q_{tx} + q_{cz} q_{tx} q_{cx} q_{cz} q_{tx})  \\
&=& \frac 14 ( 
q_{cx} + q_{cx}  q_{cz} + q_{cx} q_{tx} - q_{cx} q_{cz} q_{tx} \\
&& \hspace{15pt}- q_{cx} q_{cz}  - q_{cx} - q_{cx} q_{cz} q_{tx} + q_{cx} q_{tx}  \\
&& \hspace{15pt}+ q_{cx} q_{tx}  + q_{cx} q_{cz}q_{tx}  + q_{cx} -  q_{cx} q_{cz}  \\
&& \hspace{15pt}+ q_{cx} q_{cz} q_{tx}  + q_{cx} q_{tx}  +q_{cx} q_{cz}- q_{cx}) \\
&=& q_{cx}q_{tx} \,,
\eeas
where, the dependency on $t-1$ has again been discarded. 

The action of a gate on a descriptor can also be found directly from the matrix representation of the gate, without the detour by its functional representation and the gymnastic of the $\mathfrak{su}(2)^{\otimes n}$ algebra.
Let's exemplify the method with the case of the $\cnot$, which, in this case consists of calculating
\bes
\cnot^\dagger
	\left \{ 
		\begin{array}{c}
			\bol q_c(0) \\
			\bol q_t(0)
		\end{array}
	\right \} 
\cnot \,.
\ees
For the $q_{cx}$ element, this yields
\beas
\cnot^\dagger (\sigma_x \otimes \one) \cnot
&=&
 \begin{bmatrix}
1&0&0&0\\
0&1&0&0\\
0&0&0&1\\
0&0&1&0
\end{bmatrix} 
\begin{bmatrix}
0&0&1&0\\
0&0&0&1\\
1&0&0&0\\
0&1&0&0
\end{bmatrix}
\begin{bmatrix}
1&0&0&0\\
0&1&0&0\\
0&0&0&1\\
0&0&1&0
\end{bmatrix} \\
&=&
\begin{bmatrix}
0&0&0&1\\
0&0&1&0\\
0&1&0&0\\
1&0&0&0
\end{bmatrix}\\
&=&
\sigma_x \otimes \sigma_x \\
&=& q_{cx}(0)q_{tx}(0)\,,
\eeas
consistently with the previous approach. But why does this work?
In fact what has been computed is 
\bes
q_{cx}(1) = \Uf^\dagger_{\cnot}(\bol q(0)) q_{cx} (0) \Uf_{\cnot}(\bol q(0)) = q_{cx}(0)q_{tx}(0)\,.
\ees
The leap to the general case, \textit{i.e.}, to have $t$ and $t-1$ instead of $1$ and $0$ in the above equation, follows from observing that the calculation \emph{could have been} done by replacing~$\Uf_{\cnot}(\bol q(0))$ by its functional representation, and then use the~$\mathfrak{su}(2)^{\otimes n}$ algebraic relations at time~$0$. But since the algebraic relations are preserved,~$\bol q(0)$ could then invariably have be changed to~$\bol q(t-1)$, to obtain that, generically, $q_{cx}(t) = q_{cx}(t-1)q_{tx}(t-1)$\,.

\section{Superdense Coding, Revisited}

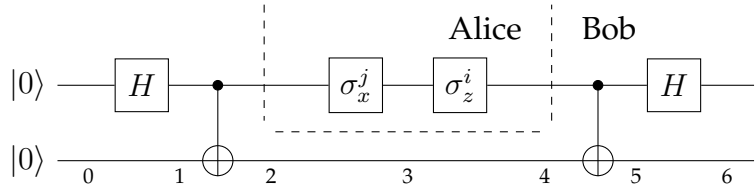
\begin{figure}[h]
	\centering
	\begin{tikzpicture}
		\node (ket1) at (0,1){$\ket 0$};
		\node (ket2) at (0,0){$\ket 0$};
		
		%H1
		\node[draw,rectangle, minimum width=0.7cm, minimum height=0.7cm
		 %right=0.8cm of ket1
		 ] (H1) at (1.5,1) {$H$};
		%CNOT1
		\node[draw, circle, minimum width=0.25cm] (T1) at (2.5,0) {};
		\filldraw (2.5,1) circle (1.8pt);
		\draw (2.5,1) --(T1.south);
				 
		 %H2
		\node[draw,rectangle, minimum width=0.7cm, minimum height=0.7cm] 
		(H2) at (8.5,1){$H$};
		%CNOT2
 		\node[draw, circle, minimum width=0.25cm] (T2) at (7.5,0) {};
		\filldraw (7.5,1) circle (1.8pt);
		\draw (7.5,1) --(T2.south);
		 
		 %Gates d'Alice, pour symétrie, placée par rapport a H1 et H2
		 \node[draw,rectangle,minimum width=0.7cm, minimum height=0.7cm, 
		 right=2.1cm of H1] (xi) {$\sigma_x^j$};
		 \node[draw,rectangle,minimum width=0.7cm, minimum height=0.7cm,
		 left=2.1cm of H2] (zj) {$\sigma_z^i$};
		 
		 %Délimitation Alice - Bob
		 \node[above left = 0.7cm and 0.7cm of xi] (p1) {};
		 \node[below left = 0.1cm and 0.7cm of xi] (p2) {};
		 \node[below right = 0.1cm and 0.7cm of zj] (p3) {};
		 \node[above right = 0.7cm and 0.7cm of zj] (p4) {};
		 \draw[dashed] (p1) -- (p2); 
		 \draw[dashed] (p2) -- (p3); 
		 \draw[dashed] (p3) -- (p4); 
		 \node[below left = 0cm and 0.1cm of p4] (alice) {Alice};
 		 \node[below right = 0cm and 0.1cm of p4] (bob) {Bob};

		%Fin
		\node (fin1) at (9.7,1){};
		\node (fin2) at (9.7,0){};
		
		%lignes
		\draw (ket1) -- (H1);
		\draw (H1) -- (xi);
		\draw (xi) -- (zj);
		\draw (zj) -- (H2);
		\draw (H2) -- (fin1);
		\draw (ket2) -- (fin2);
		
		%time labels
		\node at (0.8,-0.2) {\scriptsize{0}};
		\node at (2,-0.2) {\scriptsize{1}};
		\node at (3.2,-0.2) {\scriptsize{2}};
		\node at (5,-0.2) {\scriptsize{3}};
		\node at (6.8,-0.2) {\scriptsize{4}};
		\node at (8,-0.2) {\scriptsize{5}};
		\node at (9.2,-0.2) {\scriptsize{6}};
		%\node at (10.2,-0.2) {\scriptsize{time}};

	\end{tikzpicture}
	\caption{Network representing the superdense coding protocol.}
	\label{densecoding}
\end{figure}

In the Schr\"odinger picture, the superdense coding~\cite{bennett1992communication} may appear to hinge on 'non-local' properties of the wave-function.
% the ability for one party to `non-locally' affect the wave-function of the global system. 
See Figure~\ref{densecoding}.

The Schr\"odinger state at time~$2$ is given by the Bell state
$$\ket{\Phi^+} = \frac{\ket{00}+\ket{11}}{\sqrt 2}\,.$$
The local operations performed by Alice on her qubit shall evolve the system to one of the four Bell states in accordance with the bits~$i$ and~$j$ that she wants to transmit. The latter are then revealed by a Bell measurement. See Table~\ref{tablecoding}.

\begin{table}[h]
\centering
\begin{tabular}{|c|c|c|}
\hline
Bits $i,j$ & State at time $4$& State at time $6$\\
\hline
\hline
$0,0$ & $\ket{\Phi^+}= \frac{\ket{00}+\ket{11}}{\sqrt 2}$ \vphantom{$\frac{\int_A^A}{\int_A^A}$}& $\ket{00}= \ket{ij}$\\
\hline
$0,1$ & $\ket{\Psi^+}= \frac{\ket{01}+\ket{10}}{\sqrt 2}$  \vphantom{$\frac{\int_A^A}{\int_A^A}$}& $\ket{01}= \ket{ij}$\\
\hline
$1,0$ & $\ket{\Phi^-}=\frac{\ket{00}-\ket{11}}{\sqrt 2}$  \vphantom{$\frac{\int_A^A}{\int_A^A}$} & $\ket{10}= \ket{ij}$\\
\hline
$1,1$ & $\ket{\Psi^-}=\frac{\ket{01}-\ket{10}}{\sqrt 2}$  \vphantom{$\frac{\int_A^A}{\int_A^A}$} & $\ket{11}= \ket{ij}$\\
\hline
\end{tabular}
\caption{The Schr\"odinger state in relation to the bits to transmit.}
\label{tablecoding}
\end{table}

The protocol is now revisited in the language of descriptors. Denoting the descriptor at time~$0$ without any time labels, the computation can be done as follows.
\beas
\hphantom{\to} &\bol q (0)& \equiv 
	\left \{ 
		\hspace{-5pt}
		\begin{array}{rl}
			(q_{1x} ,& q_{1z}) \vspace{2pt}\\
			 (q_{2x} ,& q_{2z}) 
		\end{array}
		\hspace{-5pt}
	\right \} \\
\stackrel{H}{\to}& \bol q (1) & =
	\left \{
		\begin{array}{rl}
		\hspace{-5pt}
			(q_{1z},& q_{1x}) \vspace{2pt}\\
			 (q_{2x},& q_{2z}) 
		\end{array}
		\hspace{-5pt}
	\right \} \\
\stackrel{\tiny{\cnot}}{\to} & \bol q (2) & =
	\left \{
		\begin{array}{ll}
		\hspace{-5pt}
			(q_{1z}  q_{2x},& q_{1x}) \vspace{2pt}\\
			(q_{2x} ,& q_{1x} q_{2z}) 
		\end{array}
		\hspace{-5pt}
	\right \} \\
\stackrel{\sigma_x^j}{\to} & \bol q (3) & =
	\left \{
		\begin{array}{ll}
		\hspace{-5pt}
			( q_{1z}  q_{2x},& (-1)^j q_{1x}) \vspace{2pt}\\
			 (q_{2x} ,&q_{1x} q_{2z}) 
		\end{array}
		\hspace{-5pt}
	\right \} \\
\stackrel{\sigma_z^i}{\to} & \bol q (4) & =
	\left \{
		\begin{array}{ll}
		\hspace{-5pt}
			((-1)^i q_{1z}  q_{2x}, &(-1)^j  q_{1x}) \vspace{2pt}\\
			 (q_{2x} , &q_{1x} q_{2z}) 
		\end{array}
		\hspace{-5pt}
	\right \} \\
\stackrel{\cnot}{\to} & \bol q (5) & =
	\left \{
		\begin{array}{ll}
		\hspace{-5pt}
			((-1)^i q_{1z} ,& (-1)^j q_{1x}) \vspace{2pt}\\
			 (q_{2x} ,& (-1)^j q_{2z}) 
		\end{array}
		\hspace{-5pt}
	\right \} \\
\stackrel{H}{\to} & \bol q (6) & =
	\left \{
		\begin{array}{ll}
		\hspace{-5pt}
			( (-1)^j q_{1x},& (-1)^iq_{1z}   ) \vspace{2pt}\\
			 (q_{2x} ,& (-1)^j q_{2z}) 
		\end{array}
		\hspace{-5pt}
	\right \} \,.
\eeas

Denoting by~$U^{(ij)}$ the evolution throughout the the protocol,
the probability of measuring an outcome ``$i'$'' on the first qubit is given by
$$
\bra{00} U^{(ij)\dagger} (\ketbra{i'}{i'} \otimes \one) U^{(ij)} \ket{00}\,.
$$ 
In the Heisenberg picture, this computation is performed from the middle outwards.
The initial observables are expressed in terms of descriptors as
$$
\ketbra{i'}{i'} \otimes \one= \frac{\one^2 + (-1)^{i'} q_{1z}}{2} \,,
$$
which evolve by $U^{(ij)}$ to  
\bes
\frac{\one^2 + (-1)^{i'} q_{1z}(6)}{2}
= \frac{\one^2 + (-1)^{i'+i} q_{1z}}{2}  \,.
\ees
The expectation value with the reference vector~$\ket{00}$ thus yields
\bes
\frac{1 + (-1)^{i'+i}}{2}
= \delta_{ii'} \,.
\ees
Similarly, the probability of measuring ``$j'$'' on qubit 2 is given by $\delta_{jj'}$ and
hence, the system shall deterministically return the value of the bits $i$ and~$j$.

When revisited with the help of descriptors, the superdense coding of two bits into a single qubit appears quite natural: Alice's qubit's descriptor has precisely two slots in which bits can be encoded. 
When Alice transmits her qubit to Bob, measurements on that qubit alone could not leak any information about~$i$ or~$j$. In fact, any observable on Alice's qubit at time~$4$ is a linear combination of~$\one$, $q_{1x}(4)$, $q_{1y}(4) = -i q_{1x}(4)q_{1z}(4)$ and $q_{1z}(4)$, and since
\bes
\bra{00} \bol (\one, q_{1x}(4), q_{1y}(4), q_{1z}(4)) \ket{00} 
= (1,0,0,0) \,,
\ees
the expectation value of any observable on that qubit alone is independent of~$i$ and~$j$. However, the information about the bits~$i$ and~$j$ is contained in the transmitted qubit at time~$4$, since not only does~$\bol q_1(4)$ depend on~$i$ and~$j$, but those bits eventually become accessible to measurement. This kind of information, present in a system but unretrievable by measurements on the system alone has been called \emph{locally inaccessible} by Deutsch and Hayden. In step~$5$ of the protocol, Bob's qubit serves as a key as well as an extra capacity: It unlocks the bit~$i$ by getting rid of the obfuscating~$q_{2x}$ while copying the bit $j$ in its $z$ component.
 
Finally, notice that between time $2$ and time $4$, only the descriptor of the first qubit is affected, which invalidates the idea that the superdense coding protocol relies on non-local properties of entanglement. Indeed, there is an important asymmetry to be underlined: The existence of a local  way in which a phenomenon (or more generally, a theory) can be explained makes the phenomenon (or theory) local. But this doesn't hold for the attribute ``non-local'', otherwise, all phenomena and all theories would qualify as non-local by considering ad hoc non-local explanations.

\section{Conclusions}

The formalism of descriptors has been re-explained in this paper in what I hope is a more complete exposition. 
%I re-showed that this mode of description of quantum systems --- formulated in the Heisenberg picture --- is both local and complete.
I re-showed that the Heisenberg picture entails a local and complete way of describing quantum systems,
and I used the approach to revisit superdense coding. % --- locality was made explicit.
By the way, in quantum field theory, locality in the sense advocated here as no-action-at-a-distance, as well as Lorentz invariance, are also recognized in the Heisenberg picture.
%Locality, in the sense advocated here as no-action-at-a-distance, plays a central role in quantum field theory, which is also mainly formulated in the Heisenberg picture.
%I re-showed that considering quantum systems in the there exist a mode of description of quantum systems that is both local and complete. 
%
 %the descriptors were used to revisit the superdense coding entirely locally. 
 The reader who is curious to unravel the mysteries of Bell inequality violations and of quantum teleportation is referred to~\S4 and~\S5 of the article by Deutsch and Hayden (\emph{op. cit.}). 
When I explained in terms of descriptors the teleportation process to one of its pioneers, Gilles Brassard told me enthusiastically that it was the most satisfactory elucidation he had ever heard of his own invention.
%In fact, the explanation that the exposed formalism provides to the two most famous ``non-local'' manifestations of quantum theory reaches far beyond mystery breaking. 
%It roots back quantum theory together with all other scientific theories: the act of measurement needs not to be treated as a fundamentally different evolution, and it is, like any other physical process, completely local. 
%The best explanations available are unlocked by the language of descriptors, but remain oblivious in the Schr\"odinger picture. 
The best explanations of quantum processes are unlocked by the Heisenberg picture, which is manifestly local, but remain oblivious in the widespread Schr\"odinger picture.

\section*{Acknowledgements}

%, Stefan Wolf, , Philippe?. 
%Chiara, Nicetu, Sam, Lodovico

I am deeply grateful to Gilles Brassard for his benevolent support and his valuation of my research autonomy.
I am also grateful to Charles H. Bennett, Xavier Coiteux-Roy, Samuel Ducharme, Samuel Kuypers, Chiara Marletto, Pierre McKenzie, Lodovico Scarpa, William Schober and Nicetu Tibau Vidal %Jordan Payette
 %and Paul Raymond-Robichaud 
 for fruitful discussions and comments on earlier versions of this paper.
% David for important suggestions to improve the paper
% XCR, SD, PM improve the text
% Gilles et Paul, 
 I also wish to thank Stefan Wolf as well as the Institute for Quantum Optics and Quantum Information of Vienna, in particular Marcus Huber's group, for warm welcome and inspiring discussions.

This work was supported in part by the Fonds de recherche du Qu\'ebec -- Nature et technologie (FRQNT), the Swiss National Science Foundation (SNF), the National Centre for Competence in Research ``Quantum Science and Technology'' (NCCR \emph{QSIT}), the Natural Sciences and Engineering Research Council of Canada (NSERC) as well as Qu\'ebec's Institut transdisciplinaire d'information quantique (INTRIQ).

\bibliographystyle{unsrt}
\bibliography{refs.bib}

\begin{thebibliography}{1}

\bibitem{bell1964}
John~S Bell.
\newblock {On the Einstein Podolsky Rosen paradox}.
\newblock {\em Physics}, 1(3):195--200, 1964.

\bibitem{aspect1982experimental}
Alain Aspect, Philippe Grangier, and G{\'e}rard Roger.
\newblock Experimental realization of {E}instein-{P}odolsky-{R}osen-{B}ohm
  {G}edankenexperiment: a new violation of {B}ell's inequalities.
\newblock {\em Physical Review Letters}, 49(2):91, 1982.

\bibitem{bennett1993teleporting}
Charles~H Bennett, Gilles Brassard, Claude Cr{\'e}peau, Richard Jozsa, Asher
  Peres, and William~K Wootters.
\newblock {Teleporting an unknown quantum state via dual classical and
  Einstein-Podolsky-Rosen channels}.
\newblock {\em Physical Review Letters}, 70(13):1895, 1993.

\bibitem{deutsch2000information}
David Deutsch and Patrick Hayden.
\newblock Information flow in entangled quantum systems.
\newblock {\em Proceedings of the Royal Society A: Mathematical, Physical and
  Engineering Sciences}, 456(1999):1759--1774, 2000.

\bibitem{brassard2019parallel}
Gilles Brassard and Paul Raymond-Robichaud.
\newblock Parallel lives: A local-realistic interpretation of ``nonlocal''
  boxes.
\newblock {\em Entropy}, 21(1):87, authoritative version available at
  https://arxiv.org/abs/1709.10016., 2019.

\bibitem{schilppalbert1970}
Paul~A Schilpp.
\newblock {\em Albert {E}instein: philosopher-scientist}, volume~7.
\newblock The Open Court Publishing Co., 3rd revised edition, 1970.

\bibitem{gottesman1999heisenberg}
Daniel Gottesman.
\newblock The {H}eisenberg representation of quantum computers.
\newblock In {\em Group22: Proceedings of the XXII International Colloquium on
  Group Theoretical Methods in Physics}, pages 32--43. Cambridge, MA:
  International Press. \textit{Preprint quant-ph/9807006}, 1999.

\bibitem{deutsch1989quantum}
David Deutsch.
\newblock Quantum computational networks.
\newblock {\em Proceedings of the Royal Society A: Mathematical, Physical and
  Engineering Sciences}, 425(1868):73--90, 1989.

\bibitem{bennett1992communication}
Charles~H Bennett and Stephen~J Wiesner.
\newblock Communication via one-and two-particle operators on
  {E}instein-{P}odolsky-{R}osen states.
\newblock {\em Physical review letters}, 69(20):2881, 1992.

\end{thebibliography}
\end{document}